\documentstyle[preprint,eqsecnum,epsf,tighten,pra,aps]{revtex}
\begin{document}  
\title{Dimerization and Incommensurate Spiral Spin Correlations in the
Zigzag Spin Chain: Analogies to the Kondo Lattice}
\author{Steven R. White}
\address{ Department of Physics and Astronomy,
University of California,Irvine, CA 92717}
\author{Ian Affleck}
\address{Department of Physics and Canadian Institute
for Advanced Research}
\address{
University of British Columbia, Vancouver, BC, V6T 1Z1, Canada}
\date{\today} \maketitle 
\begin{abstract} Using the density matrix renormalization group and
a bosonization approach, we study a spin-1/2 antiferromagnetic
Heisenberg chain with
near-neighbor coupling $J_1$ and frustrating second-neighbor coupling $J_2$,
particularly in the limit $J_2 >> J_1$. This system exhibits both
dimerization and incommensurate spiral spin correlations.
We argue that this system is
closely related to a doped, spin-gapped phase
of the one-dimensional Kondo lattice.  
\end{abstract}

\section{Introduction}
Relatively little is known about the behavior of frustrated 
antiferromagnetic quantum spin systems, in comparison with their
unfrustrated counterparts. In general, frustration reduces antiferromagnetic
correlations and the tendency towards N\'eel order. In the presence
of sufficiently strong frustration, classical systems often develop
non-collinear sublattice magnetizations. The classical antiferromagnetic
Heisenberg triangular lattice, for example, has three different sublattices
with magnetization directions in a plane at 120$^\circ$ angles.
The behavior of the corresponding quantum system is still controversial.
Quantum systems may, instead, dimerize in
the presence of frustration, as exemplified by the exactly soluble
one-dimensional Majumdar-Ghosh model \cite{Majumdar} (see below). Yet another
possibility is some sort of spin-liquid state, without dimerization or
sublattice magnetization, examples being one-dimensional nearest
neighbor systems and possibly the Kagom\'e lattice. For a number of
reasons, frustrated systems tend to be difficult to study---for
example, quantum Monte Carlo typically cannot be used because the
frustration introduces a minus sign problem.
\begin{figure}[h]
\epsfxsize=10 cm \centerline{\epsffile{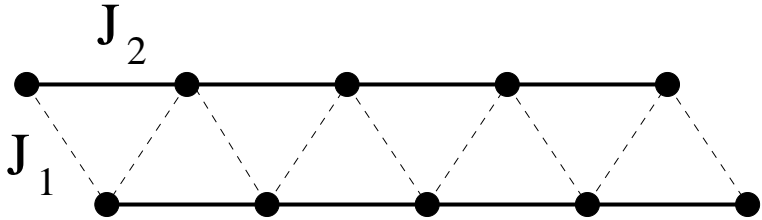}} % epsfig
\caption{The zigzag spin ladder.}
\label{fig:zigzag}
\end{figure}
In this paper, we study a one-dimensional spin-1/2 Heisenberg system, 
with near-neighbor
couplings $J_1$ and frustrating next-neighbor couplings $J_2$. 
This system can also be considered a
two-chain lattice  with diagonal, or ``zigzag'' couplings, as shown in
Figure \ref{fig:zigzag}, and we will refer to the system as the ``zigzag'' chain. The
model is described by the Hamiltonian
\begin{equation} 
H = \sum_i[J_2\vec S_i\cdot \vec S_{i+1}+J_2\vec
T_i\cdot \vec T_{i+1}+J_1\vec S_i\cdot (\vec T_i+\vec
T_{i+1})],\end{equation} 
where $\vec S_i$ and $\vec T_i$ represent the
$s=1/2$ operators on the 2 chains. 
We are particularly interested in
the case where $J_2 > J_1$. 
 
Besides a general interest
in understanding frustrated quantum systems, there are several additional
motivations for studying this system.  As we discuss below, a bosonization
treatment of a generalized version  of the one-dimensional Kondo lattice
shows a close relationship between the spin degrees of freedom of that
system when doped, and the zigzag chain. The evidence we present below for
a spin gap in the zigzag chain may also indicate a spin-gapped phase in a
Kondo chain. Physically, zigzag arrangements of atoms are common and 
 this model is appropriate for  real systems such as SrCuO$_2$,
studied in Ref. \onlinecite{Matsuda}. 

\begin{figure}[h]
\epsfxsize=10 cm \centerline{\epsffile{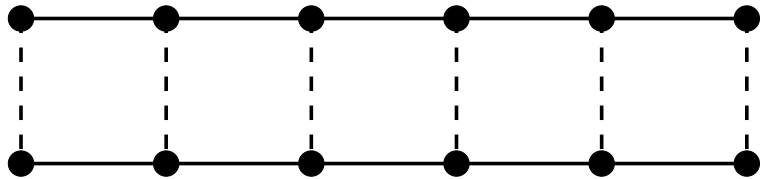}} % epsfig
\caption{The spin ladder.}
\label{fig:spin_ladder}
\end{figure}
This model
exhibits some similarities, but also marked differences from the standard
two-chain spin-ladder\cite{dagotto,strong,barnes,rice,rvbprl}, 
shown in Figure \ref{fig:spin_ladder}. While inter-chain coupling is
relevant in the ladder model, producing a gap which scales linearly (up to
logarithmic corrections) for either sign of the coupling, it is marginal
for the zigzag chain, producing a gap only for antiferromagnetic sign,
which scales exponentially with coupling, 
$ \Delta \propto \exp(-\hbox{ constant }J_1/J_2)$. 
The marginal nature of the inter-chain
coupling in our model, is reminiscent of weakly coupled two-chain Hubbard
systems\cite{Schulz1,Balents}.  However, the renormalization group
analysis of our model is both simplified by the absence of charge
excitations and complicated by the presence of irrelevant intra-chain
operators with  coupling constants of $O(1)$.
Unlike in the spin-ladder model, our numerical results for the
spin-spin correlation functions indicate that dimerization is
present for all $J_1<J_{1c}\approx 4 J_2$ and
incommensurate spiral correlations are present for all $J_1< 2 J_2$. 

From the numerical point of view, for a 1D or quasi-1D system,
this model is quite challenging. As mentioned above, quantum Monte Carlo
methods are not useful. Because of the very long correlation lengths, exact
diagonalization is not useful for the range of parameters in which
we are interested. 
The model can be studied with the density matrix renormalization
group\cite{dmrg} (DMRG), but it is still a difficult system for several
reasons. The exceptionally long correlation lengths mean that very long
systems must be studied---in some cases we have studied systems with
thousands of sites. Furthermore, the system, particularly for large $J_2$,
has two properties which slow the convergence of DMRG with the number of
states kept per block: 1) the correlation
length is long, and 2) it is composed of nearly {\it independent } 
chains. The slower convergence of DMRG with nearly independent chains
is known from the two-chain spin ladder case.
For $J_2/J_1 = 2$, for example, it was necessary to keep about 400 
states per block for an adequate calculation of the spin-spin correlation 
function. In contrast, similar accuracy can be obtained in the $S=1$
Heisenberg chain keeping 50 to 80 states. The calculations presented
here were only possible because of new developments in the DMRG algorithms,
resulting in increased computational capabilities by nearly two orders
of magnitude, which will be mentioned here but discussed in more 
depth elsewhere.

Previous studies of this system include the exact diagonalization
work of Tonegawa and Harada\cite{tonegawa}, the recent work
of Bursill, et. al. \cite{bursill} using DMRG and a coupled
cluster method, and the DMRG study of Chitra, et. al. \cite{chitra}, who
considered a more general model which included a dimerization term.
This work extends and improves upon the previous work in several
ways.  Particularly for the difficult $J_2 > J_1$ regime, we have been 
able to obtain improved numerical results, which are consistent with
the previous calculations.  We have presented the first field theory
treatment of the zigzag chain in the $J_2>>J_1$ regime, as far as we
know, and found general agreement with the numerical results.  In
addition, we point out a potentially important connection between
the zigzag chain and the generalized Kondo lattice.

This paper is organized as follows. In Section II, we discuss a bosonization
and renormalization group approach for the zigzag chain. In Section
III, we discuss the DMRG methods used in the numerical calculations.
In Section IV, we
present numerical results for a variety of  properties.  In Section V
we discuss the relationship of the spin ladder and the zigzag
chain to the half-filled and non-half-filled Kondo lattice,
respectively.  Section VI contains conclusions.

\section{Bosonization and Renormalization Group treatment of the Zig-Zag
Spin Chain}

Quite different field theory treatments of the zigzag spin chain are
appropriate depending on the ratio $J_1/J_2$.  For small values of
$J_2/J_1$ it is appropriate to treat the model as a single chain.
Extensive discussion of this model may be found in
Refs.\onlinecite{Haldane,Affleck1,Eggert}. The conclusion is that
the model is gapless for $J_2\leq J_{2c}\approx
0.241167$\cite{Eggert2,okamoto}. We
briefly review the conclusions here.  The low energy effective
theory is a free massless boson with SU(2) symmetry, or equivalently
a level k=1 Wess-Zumino-Witten (WZW)  non-linear $\sigma$-model.
The uniform and staggered components of the spin operators are
represented in terms of the left and right-moving density operators
$\vec \rho_{L,R}$ and the $SU(2)$ matrix field $g$ of the WZW model as: 
\begin{equation}
\vec S_j \approx (\vec \rho_{L}+\vec \rho_{R})
+\hbox{constant}(-1)^j \hbox{tr}(\vec \sigma g).\label{bosonize}
\end{equation} 

\begin{figure}[h]
\epsfxsize=10 cm \centerline{\epsffile{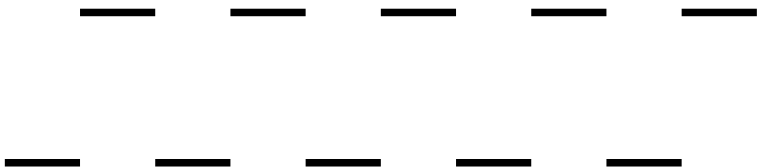}} % epsfig
\caption{The two dimer groundstates of the Majumdar-Ghosh model.}
\label{fig:MG}
\end{figure}

As well as the WZW or free boson
Hamiltonian, there is an additional marginal interaction controlled by $J_2$:
\begin{equation} H_{\hbox{int}}\propto (J_2-J_{2c})\vec \rho_L\cdot \vec
\rho_R. \label{Hintu}\end{equation}
This is marginally irrelevant for $J_2<J_{2c}$; that is it renormalizes to
zero at long length scales but produces logarithmic corrections to the
simple scaling behavior of the free boson model.  For $J_2>J_{2c}$ it is
marginally relevant; that is it renormalizes to large values producing an
exponentially small gap and inverse correlation length:
\begin{equation} \Delta =v_s/\xi \propto
e^{-\hbox{constant}J_1/(J_2-J_{2c})}. 
\end{equation} 
This massive phase is spontaneously dimerized.  The order parameter:
\begin{equation} d = <\vec S_{2i-1}\cdot \vec S_{2i}>-
<\vec S_{2i}\cdot \vec S_{2i+1}>\propto <\hbox{tr} g>.
\end{equation}
Since tr $g$ has scaling dimension $1/2$ it follows that:
\begin{equation} d\propto \sqrt{\Delta}\end{equation}
As $J_2/J_1$ is increased further the correlation length decreases and the
field theory description becomes less accurate.  At the point $J_2/J_1=.5$
the exact groundstates become the simple dimer configurations of Figure 
\ref{fig:MG}, 
as first pointed out by Majumdar and Ghosh \cite{Majumdar}.  Here the
correlation length is only one lattice spacing; a rigorous proof of a gap
has been given in this case \cite{Affleck}.

As $J_2/J_1$ is increased
further, we expect the correlation length to increase again.  Eventually,
when $J_2>>J_1$, it is more appropriate to think of two spin chains with a
weak zigzag inter-chain coupling, as shown in Figure \ref{fig:zigzag}.  At $J_1=0$, we
obtain two decoupled chains with vanishing gap and inverse correlation
length.  For small $J_1/J_2$ we expect the gap and inverse correlation
length to be small so we may perturb around the point $J_1=0$ using a field
theory approximation.  Only the low energy degrees of freedom of each chain
are relevant.  We may represent these as in Eq. (\ref{bosonize}),
introducing two sets of fields $g_i$, $\vec \rho_{L,i}$ and $\vec
\rho_{R,i}$ where $i=1,2$ labels the two chains.  The difference between the
ladder and the zigzag chain becomes evident upon bosonizing the
inter-chain coupling.

In the ladder model we obtain a coupling of the alternating spin components:
\begin{equation} H_{\hbox{int}a} \propto J_2 \hbox{tr}(\vec \sigma g_1)\cdot 
\hbox{tr}(\vec \sigma g_2).\end{equation}  This has dimension 1, so we
expect it to produce a gap proportional to $J_2$ (up to logarithmic
corrections associated with the marginal operators).  To check explicitly
that all degrees of freedom are gapped it is convenient to switch over to
abelian bosonization notation. The SU(2) fields $g_i$ can be written in
terms of the free spin bosons, $\phi_{si}$ as: \begin{equation} 
g_j \propto \left(\begin{array}{cc}e^{i\sqrt{2\pi}\phi_{sj}} & 
e^{i\sqrt{2\pi}\tilde \phi_{sj}}\\
-e^{i\sqrt{2\pi}\tilde \phi_{sj}} & e^{-i\sqrt{2\pi}\phi_{sj}}
\end{array}\right).\label{gdef}\end{equation}
Here $\tilde \phi$ is the field dual to $\phi$
[$\int_{-\infty}^x dx' \Pi (x')$].  [The subscript $s$ for spin is
redundant here but we keep it to distinguish these fields from the
charge boson that arises in the discussion of the Kondo lattice in
Sec. V.]  Thus the alternating spin
operators are: \begin{equation} \hbox{tr}(\vec \sigma g_i) \propto
(\sin \sqrt{2\pi}\tilde \phi_{si} ,-\cos \sqrt{2\pi} \tilde \phi_{si} ,
\sin \sqrt{2\pi}\phi_{si} ).\label{abelian}\end{equation}  Introducing the
sum and difference spin bosons: \begin{equation}\phi_{s,\pm}\equiv
{\phi_{s1}\pm \phi_{s2}\over \sqrt{2}},\end{equation}
$H_{\hbox{int}a}$ becomes: \begin{equation} H_{\hbox{int}a}\propto
-\cos (\sqrt{4\pi}\phi_{s+}) +\cos
(\sqrt{4\pi} \phi_{s-})+(1/2)\cos (\sqrt{4\pi}\tilde \phi_{s-}).
\label{abelint}\end{equation} Thus we obtain decoupled Hamiltonians for the
two fields $\phi_{s\pm}$.  The interactions are expected to produce
gaps for both bosons for either sign of $J_2$, proportional to
$|J_2|$.\cite{Schulz2,Strong,Shelton}

\begin{figure}[h]
\epsfxsize=12 cm \centerline{\epsffile{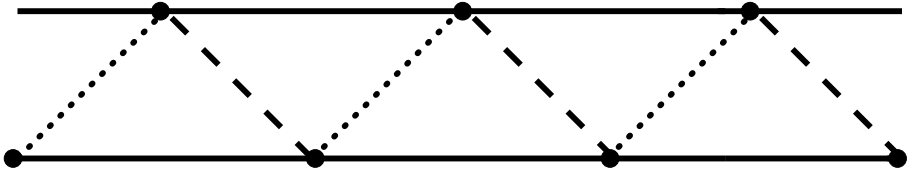}} % epsfig
\caption{Dimerization in the zigzag spin chain.}
\label{fig:zigzagdimer}
\end{figure}

In the zigzag model the coupling of alternating spin components
cancels, and we are left only with the marginal coupling of uniform
components:
\begin{equation} 
H_{\hbox{int}u}=2J_1(\vec \rho_{L1}+\vec \rho_{R1})\cdot
(\vec \rho_{L2}+\vec \rho_{R2}).\end{equation}
The term $(\vec \rho_{L1}\cdot \vec \rho_{L2}+\vec \rho_{R1}\cdot
\vec \rho_{R2})$ does not renormalize to lowest order; we will assume it
can be ignored, apart from a small velocity renormalization.
The remaining Hamiltonian is Lorentz
invariant. Note that the unperturbed spin Hamiltonian separates into 4
terms for left and right moving spin bosons of type 1 and 2.  Thus we may
interchange $\vec \rho_{R1}$ with $\vec \rho_{R2}$, so that the interaction
becomes diagonal in the index i, labeling the two species of spin bosons. 
This interaction is then precisely two copies of the one which occurs for
a single s=1/2 chain, Eq. (\ref{Hintu}). [This decoupling is spoiled by
irrelevant operators.] Thus we conclude that, for ferromagnetic zigzag
coupling, the interaction renormalizes logarithmically to zero and there
is no gap.  For antiferromagnetic coupling there is an exponentially
small gap: 
\begin{equation} \Delta = v_s/\xi \propto
e^{-\hbox{constant}J_2/J_1}.
\end{equation}
We might again expect a broken discrete translational symmetry.  Since
the marginally relevant interaction couples left and right movers on the
two chain, it is natural to expect that the order parameter is:
\begin{equation}
d\equiv <\vec S_{2i}\cdot(\vec T_{2i}-\vec T_{2i+1})>,
\label{dimerdef}\end{equation} 
as shown in Figure \ref{fig:zigzagdimer}.  Note that, when we regard
the model as a single chain with first and second nearest neighbor
interactions, this is the same nearest neighbor dimer order parameter
that was discussed above.  Thus there is need be no other phase transition
for $J_{2c}\approx 0.25J_1<J_2<\infty$.  In two-chain field theory language,
the order parameter is $\hbox{tr} (\vec \sigma g_1)\cdot \hbox{tr} (\vec
\sigma g_2)$.  This has dimension 1, so it should scale linearly with
$\Delta$
\begin{equation} d \propto \Delta  
\propto e^{-\hbox{constant}J_2/J_1},\label{dexp}
\end{equation}
 unlike in the other limit, discussed above where
 $d\propto \sqrt{\Delta}$.  Note that, in this
phase, the much stronger intra-chain correlations, $<\vec T_i\cdot \vec
T_{i+1}>$ are translationally invariant; only the very weak inter-chain
correlations break translational symmetry.  

In addition to the long-range dimer order, there is also finite-range
N\'eel order, with correlation length $\xi$.  $\xi$ becomes very long
when $J_2$ is only slightly larger than $J_{2c}\approx 0.25$ and again
when $J_2>>J_1$.  Near $J_{2c}$ this long but finite range order is of
standard N\'eel type:
\begin{equation} <\vec S_i\cdot \vec S_j>\propto
(-1)^{i-j}e^{-|i-j|/\xi}.\end{equation}  In the other limit, $J_2>>J_1$, 
the spins on each chain exhibit finite range N\'eel order:
\begin{equation} <\vec S_i\cdot \vec S_j>\propto <\vec T_i\cdot \vec T_j>
\propto (-1)^{i-j}e^{-|i-j|/\xi}.\end{equation}
Note that from the single chain point of view, it is the second nearest
neighbor spins which are N\'eel ordering.  

Further insight into the behavior of this model can be obtained from 
solving the classical problem.  This gives a spiral order with pitch
angle $\theta$.  $\theta$ is the angle between neighboring spins in the
single chain description.  
The classical energy per unit length is:
\begin{equation} 
E(\theta ) =J_1\cos \theta + J_2\cos 2\theta .\end{equation}
For $J_2\leq 0.25 J_1$ the solution is the standard N\'eel state, but for
$J_2>J_1$ the pitch angle is given by:
\begin{equation} \cos \theta = -J_1/4J_2.\label{angle}\end{equation}
This angle varies from $\pi$ at $J_2\to 0.25 J_1$ to $\pi /2$ at
$J_2/J_1\to \infty$.    Of course, true N\'eel order is presumably
impossible in a one-dimensional quantum antiferromagnet, even at zero
temperature.  However, we might expect the classical result to give a
guide to possible finite-range order. 
Note that, in the case $J_2>>J_1$, the spins on
different chains are almost decoupled, $|<\vec S_i\cdot \vec T_i>|<<1$,
so classically we may regard these spins as orthogonal.  This is
consistent with the $\pi /2$ pitch angle predicted classically.  
We expect that the deviation of the pitch
angle from $\pi /2$ defines a characteristic wave-length, which
we expect to be proportional to $\xi$: i.e.
\begin{equation} 
\theta - \pi /2 \propto 1/\xi \ \ (J_2>>J_1).
\end{equation}
For smaller values of $J_2/J_1$, Chitra, et. al.\cite{chitra} showed
(using DMRG) that the spiral correlations are only present for 
$J_2/J_1 > 0.5$, in agreement with the earlier work of Tonegawa and
Harada\cite{tonegawa}. Between the critical point and the
Majumdar-Ghosh point, the pitch angle is $\pi$.

\section{The Density Matrix Renormalization Group approach}
The DMRG method is discussed in some detail in references \onlinecite{dmrg}
and \onlinecite{georgia}. Here we discuss specifics of our calculations,
as well as mentioning some recent improvements to DMRG which are
implemented in our calculations.

The numerical results were based on finite-system studies of systems
with as many as 600 sites, and on infinite-system results for systems
up to 7000 sites.  We obtain the total energies, bond energies,
and the equal--time spin-spin correlation functions of the
ground state.
The accuracy of the calculations depend on the number of states $m$
kept per block, as well as the value of $J_2/J_1$. In these calculations
we used values of $m$ up to 700. Large values of $m$ were necessary
only for large values of $J_2/J_1$. Near the Majumdar-Ghosh point,
$J_2/J_1=0.5$, very small values of $m$ sufficed; exactly at the 
Majumdar-Ghosh point, exact results for the ground state could be 
obtained with $m=2$.
Truncation errors, given by the sum of the density matrix
eigenvalues of the discarded states, ranged from zero to 
O($10^{-8}$) for $J_2/J_1=2.5$.
This discarded density matrix weight is directly correlated with the
absolute error in the energy \cite{dmrg}.
We apply open boundary conditions to the lattice because
the DMRG method is most accurate for a given amount of computational
effort with these boundary conditions.

We have been able to perform much more extensive calculations, both
in terms of system size and number of states kept, than in previous
DMRG studies. This is not because of more extensive computational
resources--these calculations were performed almost entirely on
a Digital AlphaStation 200 4/166 with 96 megabytes, rated at 
135 SPECfp 92. The primary reason for the improved capabilities are
some recent improvements to the DMRG algorithms. The main improvement
involves a transformation of the wavefunction from the previous
DMRG step to the current DMRG step, providing an excellent starting
point for the sparse matrix diagonalization procedure. These improvements
will be reported elsewhere \cite{improvedmrg}. In addition, the
calculations were performed using a highly optimized C++ program,
which, for example, translates all matrix operations into calls to
a very efficient Basic Linear Algebra Subroutine (BLAS) library.
This program keeps in memory only what is necessary at each step;
everything else is written to disk.

\section{Results}
\begin{figure}[h]
\epsfxsize=10 cm \centerline{\epsffile{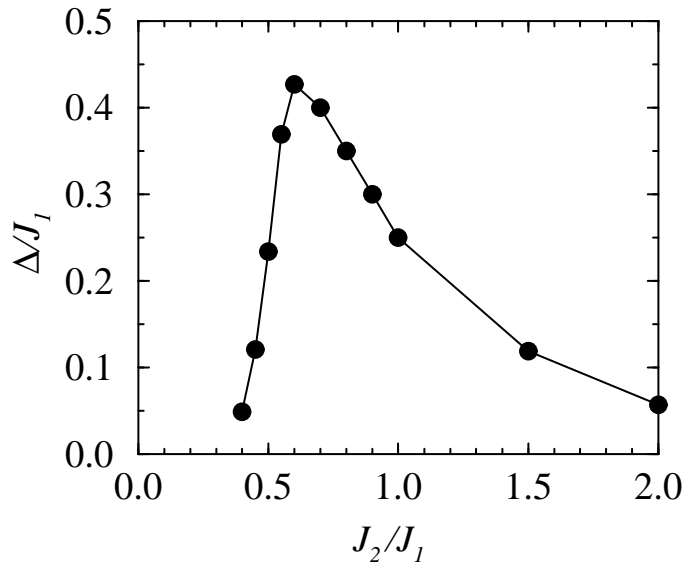}} % epsfig
\caption{The spin gap.}
\label{fig:spin_gap}
\end{figure}

The spin gap was calculated for $J_2/J_1$ ranging from 0.4 to 2.0. 
The gap is defined as the difference in energy between the lowest
$S_z=0$ state and the lowest $S_z=1$ state. For each value of $J_2/J_1$,
this energy difference was calculated for a set of finite systems with
open boundary conditions with different sizes $L$, with $L$ typically
ranging from 32 to 200. (We consider the system to be a $L\times 2$
lattice, so the number of sites is actually $2L$.) The gaps were extrapolated
to $L \to \infty$ using a polynomial fit of the form
\begin{eqnarray}
\Delta_L = \Delta + a_2/L^2 + a_3/ L^3 + \ldots,
\end{eqnarray}
and also including an $a_1/L$ term. 
The extrapolation was much more problematic than, for example, in
the Heisenberg $S=1$ chain, where the $a_1/L$ term should be
omitted \cite{boseconden}. For the zigzag chain, the $a_1/L$ term appeared
to be the {\it dominant} behavior except for exceptionally long systems.
We dealt with this problem by using very long chains, so that the
extrapolation itself was small.
Our results are shown in Figure \ref{fig:spin_gap}.

\begin{figure}[h]
\epsfxsize=10 cm \centerline{\epsffile{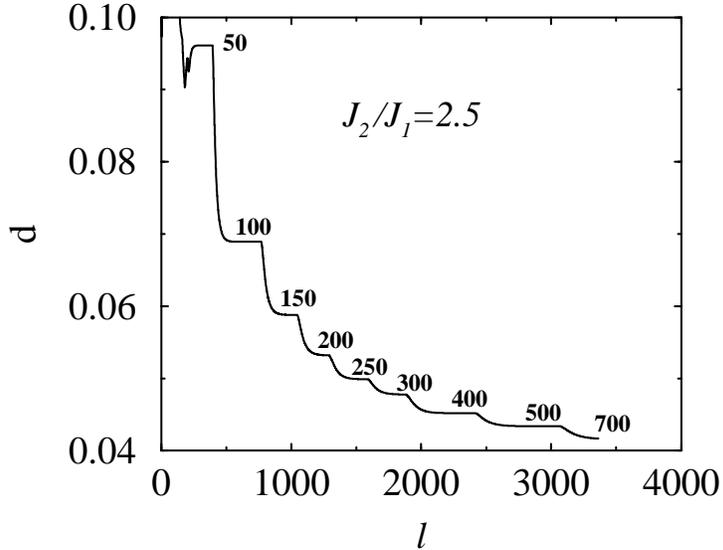}} % epsfig
\caption{The dimerization during an infinite system DMRG calculation
as a function of the DMRG step $l$. The numbers labeling each plateau
are the number of states kept per block; once convergence was reached,
the program was signaled to keep additional states.}
\label{fig:dimerml}
\end{figure}

\begin{figure}[h]
\epsfxsize=10 cm \centerline{\epsffile{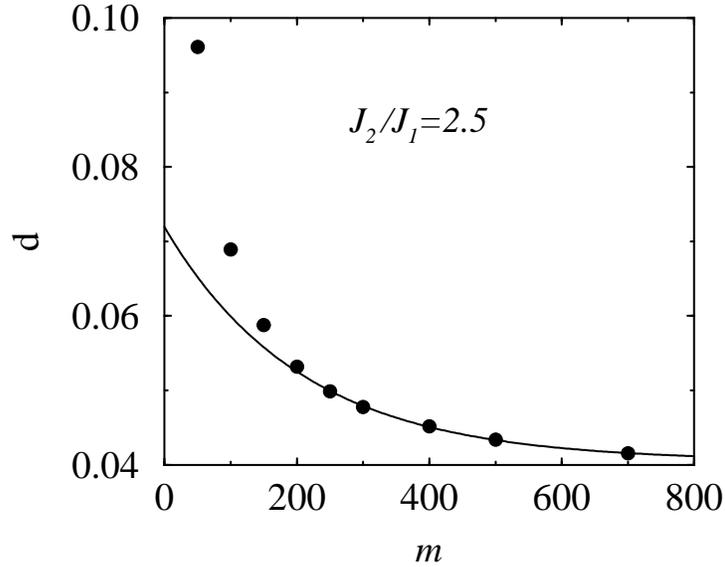}} % epsfig
\caption{The dimerization plateau values from Figure 6. 
The solid line is a fit of the form
$y=0.0405 + 0.03144 \exp(-m/208)$.}
\label{fig:dimerm}
\end{figure}

\begin{figure}[h]
\epsfxsize=10 cm \centerline{\epsffile{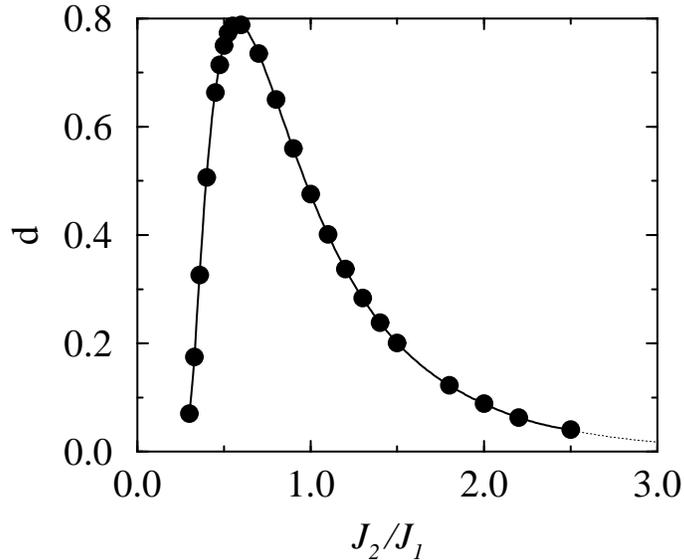}} % epsfig
\caption{The dimerization as a function of $J_2/J_1$. The solid line
is a spline fit to the data, and the dotted line for $J_2/J_1 > 2.5$ is
an exponential fit $d = 2.283\exp(-1.622 J_2/J_1)$.}
\label{fig:dimer}
\end{figure}

We define the dimerization to be the absolute value of the 
difference of adjacent interchain bond strengths, given explicitly
by Eq. (\ref{dimerdef}).
We calculated the dimerization using the infinite system DMRG
method, extrapolating in the number of states kept per block $m$. 
The ground state is doubly degenerate, corresponding to a shift
in the dimerization by one site. Ordinarily we expect to get
a ground state with a fully broken dimerization symmetry, but
it is possible numerical effects could produce an intermediate
value for the dimerization. To prevent this, we add a very
small dimerization field, proportional to Eq. (\ref{dimerdef}),
to the Hamiltonian, with the coefficient taken as $10^{-5}$.
A very long calculation was performed, starting with a relatively
small $m$. Once the dimerization converged with system size 
for this value of $m$, $m$ was increased, and allowed to converge again.
Figure \ref{fig:dimerml} shows the most difficult case we studied,
 $J_2/J_1=2.5$. 
A maximum of $m=700$ was used in this calculation, but substantially
smaller values of $m$ were necessary for most values of $J_2/J_1$.
The plateau values are shown as a function of $m$ in Figure \ref{fig:dimerm}. 
These can be fit very well with an exponential form, allowing the extrapolation
to $m \to \infty$. (Near the Majumdar-Ghosh point, this procedure was
not necessary: nearly exact results were easily obtained using small
values of $m$.) The final results, corresponding to $m \to \infty$ and
$L \to \infty$, are shown in Figure \ref{fig:dimer} . 
One interesting feature of these
results is that the maximum dimerization does not occur at the 
Majumdar-Ghosh point, which one normally considers to be fully
dimerized, with dimerization $3/4$. 
A maximum of $0.7906135$ occurs at approximately $J_2/J_1 = 0.5781$.
(For the calculation of the maximum we reduced the value of the
dimerization field to $10^{-8}$.)
A dimerization greater than $3/4$ is possible if
one of the bonds is slightly {\it ferromagnetic}.
Consequently, a maximum dimerization of unity would be possible if
the two bonds were independent. Since they are not independent,
the maximum possible dimerization is reduced. We can obtain the
maximum possible dimerization by considering a model with the dimerization
order parameter as its Hamiltonian
\begin{equation}
H = \sum_i\vec S_{2i}\cdot(\vec T_{2i}-\vec T_{2i+1}).
\end{equation} 
The ground state of this system has the maximum
possible dimerization. DMRG converges very rapidly for this Hamiltonian,
yielding a dimerization of $0.8245165$.
Returning to the original zigzag Hamiltonian, we find that
the weaker bond is ferromagnetic over a broad region: from 
$0.5 < J_2/J_1 < 2.5$. The upper limit is somewhat uncertain,
since we do not have reliable results for $J_2/J_1 > 2.5$.
For large values of
$J_2/J_1$, the dimerization can be fit very well by the function $d =
2.283\exp(-1.622 J_2/J_1)$, as predicted in Eq. (\ref{dexp}), which is shown as the dotted line in the
figure.

\begin{figure}[h]
\epsfxsize=11 cm \centerline{\epsffile{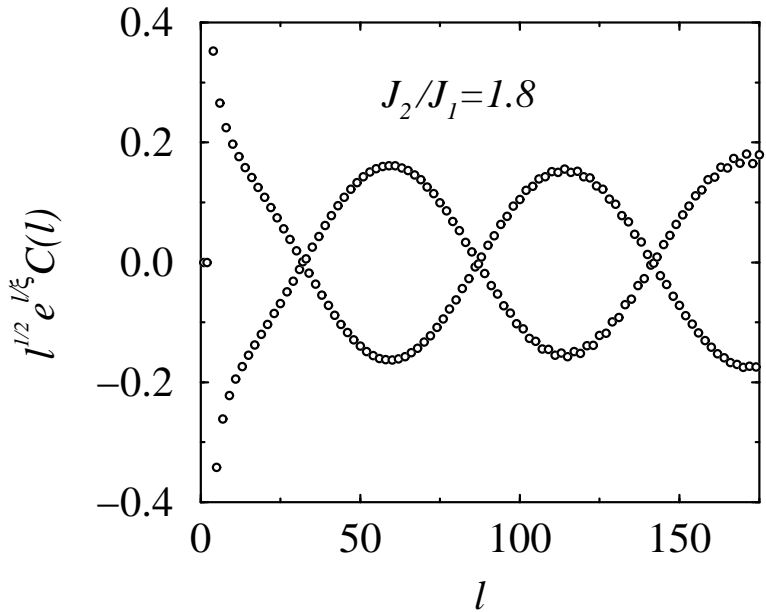}} % epsfig
\caption{The correlation function $C(l) = \langle \vec S \cdot \vec S
\rangle$, multiplied by $l^{1/2}e^{l/\xi}$, where the correlation length
is $\xi = 17.1$. The system size was $200\times 2$, and $m=350$ states
were kept per block.} \label{fig:corlenl}
\end{figure}

The spin-spin correlation function 
$C(i-j) = \langle \vec S_i \cdot \vec S_j\rangle$
was calculated as a function of $i-j$
using the finite system method, with $i$ and $j$ chosen
symmetrically about the center of the system.
The correlation function exhibits incommensurate behavior. This
behavior is clearly seen in Figure \ref{fig:corlenl}, where we plot $C(l)$ divided by the
asymptotic form of the Lorentz invariant free massive boson propagator,
to eliminate the dominant decay behavior of the function. 
The correlation length $\xi$ was chosen to make the maximum in
the ``beat'' amplitudes as constant as possible.  The equal-time free
boson propagator is:
\begin{equation} \int {d\omega dk \over (2\pi )^2}{e^{ikl}\over
\omega^2/v^2+k^2+\xi^{-2}}\propto \int dk{e^{ikl}\over
\sqrt{k^2+\xi^{-2}}}  \ \ { _{\longrightarrow} \atop ^{l>>\xi}}\ \   
l^{-1/2}e^{-l/\xi}.
\end{equation}
The fit was 
also performed without the factor of $l^{1/2}$, with noticeably 
poorer results. Our results for the correlation length are subject
to greater errors than, for example, the dimerization, both because
a precise fitting function is not known and the errors in correlation
functions are greater than in local measurements in DMRG calculations.

\begin{figure}[h]
\epsfxsize=11 cm \centerline{\epsffile{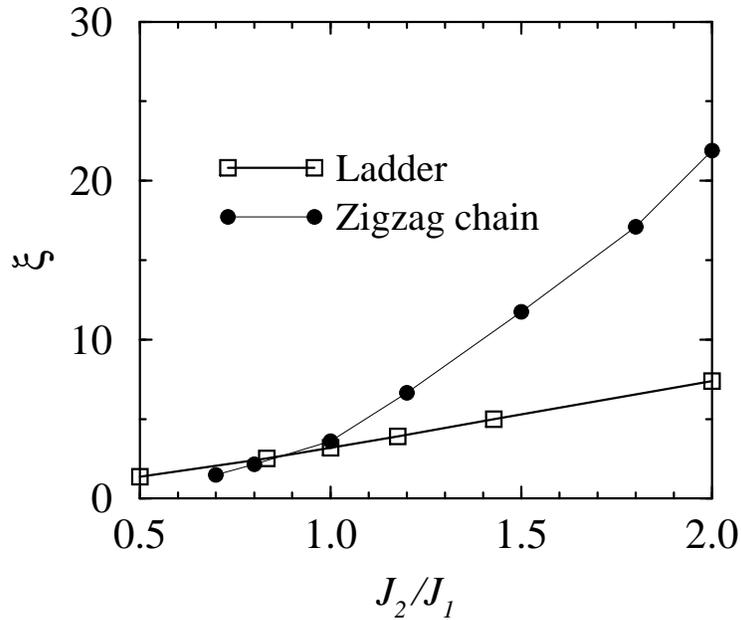}} % epsfig
\caption{The correlation length as a function of $J_2/J_1$.}
\label{fig:corlen}
\end{figure}

\begin{figure}[h]
\epsfxsize=11 cm \centerline{\epsffile{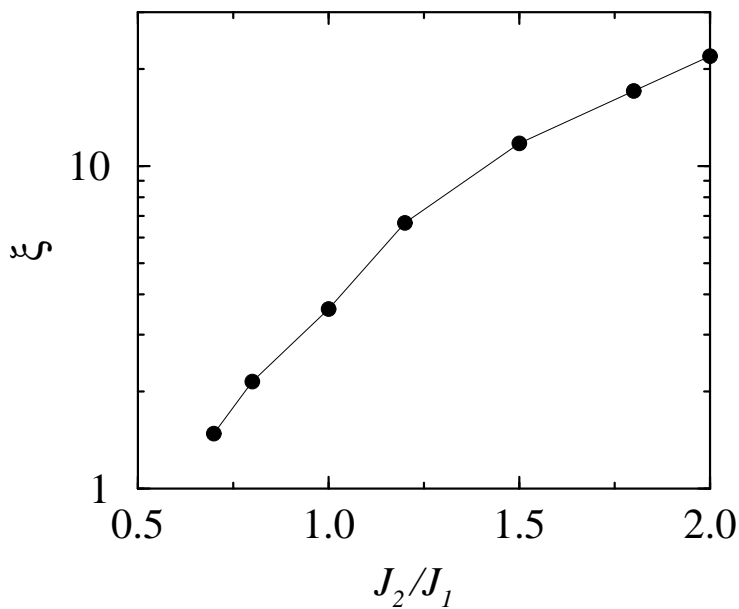}} % epsfig
\caption{A semilog plot of the correlation length as a function of $J_2/J_1$.}
\label{fig:logxi}
\end{figure}

The resulting values for $\xi$ are shown in Figure \ref{fig:corlen}. For comparison
we show similar results for the ladder system shown in Fig. 2.
The correlation
length grows much more rapidly in the zigzag chain than in the ladder
system. (These lengths correspond to considering the zigzag chain
a $L\times 2$ system. If we consider it a single chain, these lengths
would be multiplied by two.) Figure \ref{fig:logxi} shows the same 
results as a semilog plot. It appears that for the accessible
values of $J_2/J_1$, the correlation length does not yet grow exponentially
with $J_2/J_1$.

\begin{figure}[p]
\epsfxsize=9 cm \centerline{\epsffile{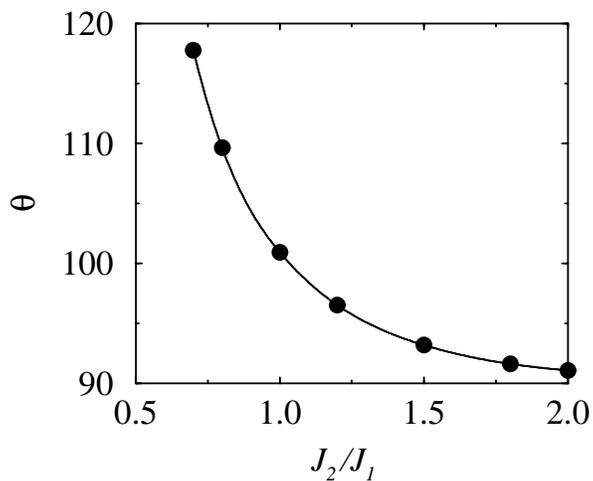}} % epsfig
\caption{The angle $\theta$ versus $J_2/J_1$.}
\label{fig:angle}
\end{figure}
\begin{figure}[p]
%\parbox{6in}{
\epsfxsize=9 cm \centerline{\epsffile{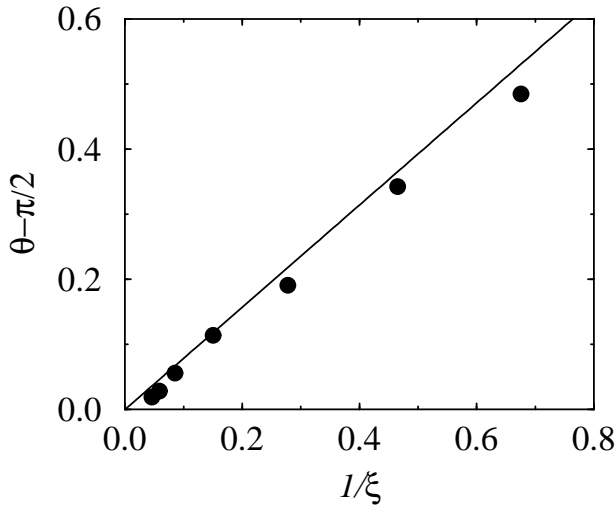}} % epsfig
\caption{ $\theta -\pi /2$ versus $1/\xi$.}
%}
\label{fig:anglexi}
\end{figure}

By fitting $l^{1/2}e^{l/\xi}C(l)$ to a sinusoidal function with
an arbitrary wavelength and phase, we were able to extract the 
incommensurate angle $\theta$, which is shown in Figure \ref{fig:angle}. 
In Figure \ref{fig:anglexi}, we plot $\theta-\pi/2$ versus $1/\xi$, for which
we expect linear behavior near the origin. 
The solid line corresponds to 
\begin{eqnarray}
(2\xi)^{-1} = \frac{\theta}{\pi/2} - 1.
\end{eqnarray}
Given the uncertainties in our procedures for extracting $\theta$ and $\xi$,
our data points are reasonably consistent with this behavior.

\section{Kondo Lattice}
We consider a generalized one-dimensional Kondo lattice Hamiltonian:
\begin{equation}
H=\sum_i[-t(c^\dagger_{i}c_{i+1}+\hbox{h.c.})+J \vec
S_i\cdot c^\dagger_i{\vec \sigma \over 2}c_i + K\vec S_i\cdot \vec
S_{i+1}].\end{equation}
Here $c_{i,\sigma}$ anihilates an electron of spin $\sigma$ at site
$i$ and $\vec S_i$ is a spin-1/2 operator.  Sums over spin indices
are implicit.  We include an explicit nearest neighbour spin coupling,
$K$.  This (and longer range terms) would be generated by conduction
electron exchange (the RKKY mechanism).  It could also arise from
other exchange mechanisms in some cases as for instance in the
organic chain compound CuPC(I).\cite{Ogawa,Quirion}
We include it here because it
allows for straightforward application of bosonization techniques. 

In fact the limit to which our method applies is $J <<t,K$.  In
this limit, we are only concerned with the low energy degrees of
freedom of the conduction electrons ($c_i$) and the spin chain.
These can be represented in bosonized form, introducing spin ($\phi_{s1}$)
and charge ($\phi_c$) bosons to represent the conduction electrons and
an additional spin boson ($\phi_{s2}$) to represent the spin chain. 
This approach was attempted independently in Ref.
\onlinecite{Fujimoto} but we disagree with their conclusions as
explained below.  The
two spins bosons turn out to be very similar to the ones discussed for
two spin chains above.  They
spin bosons may equivalently be represented by the matrix fields $g_i$.
For $J=0$, we obtain simply 3 decoupled free boson Hamiltonians (plus 
various irrelevant interactions involving $\phi_{s2}$).  The velocities
are $v_F=2t$ for $\phi_c$ and $\phi_{s1}$ and $v_s=\pi K/2$ for
$\phi_{s2}$.  We note that the theory is not Lorentz invariant due to the
difference of velocities.
 To bosonize the Kondo interaction we need the bosonic representations
for the conduction electron spin operator and the localized spin
operators, $\vec S_i$.  These are given by: \begin{eqnarray} 
 c^\dagger_j{\vec \sigma \over 2}c_j&\approx& (\vec \rho_{L1}+\vec
\rho_{R1}) +\hbox{constant}\cdot [e^{2ik_Fj}\hbox{tr}(\vec \sigma
g_1)e^{i\sqrt{2\pi}\phi_c}+\hbox{h.c.}]\nonumber \\
\vec S_j &\approx& (\vec \rho_{L2}+\vec
\rho_{R2}) +\hbox{constant}(-1)^j \hbox{tr}(\vec \sigma
g_2).\label{bosonizeK}
\end{eqnarray}  
Note that, away from
half-filling, the continuum limit Hamiltonian only contains a
marginal coupling of the currents:
\begin{equation}H_{\hbox{int}u} = J(\vec \rho_{L1}+\vec
\rho_{R1})\cdot (\vec \rho_{L2}+\vec \rho_{R2}).\end{equation}  
At half-filling, the $2k_F$
oscillation becomes commensurate with the alternating localized spin
operator and an additional relevant (dimension 3/2) term occurs:
\begin{equation}
H_{\hbox{int}a} = \lambda \hbox{tr}(\vec \sigma g_1)\cdot 
\hbox{tr}(\vec \sigma g_2)\cos (\sqrt{2\pi}\phi_c),\end{equation}
where $\lambda \propto J$.

The strong analogy with the two-chain spin system is now clear.  The
half-filled Kondo lattice  resembles the standard spin ladder
model, with a relevant inter-chain interaction.  Away from half-filling
the Kondo lattice is very closely analogous to the zigzag spin chain
model.  In this case there are no relevant charge interactions so we may
expect a decoupled massless charge sector.  The field theory describing
the spin sector is identical to that occuring in the zigzag spin chain,
except for the difference of spin-wave velocities for the two chains.  In
fact, this situation could also easily be realised in the zigzag spin
chain by having different coupling constants $J_2$ and $J_2'$ for the two
chains.  It is also clear that adding Hubbard interactions for the
electrons does not change things very much, particularly away from
half-filling.  The charge excitations still decouple and remain gapless
in that case.  The scaling dimensions of various operators change with
the Hubbard interaction strength, associated with rescaling the charge
boson, in the standard way.

At half-filling we expect $H_{\hbox{int}a}$, which couples charge and
spin modes, to produce a gap for all charge and spin excitations. The gaps
should scale as $J^2$ for either sign of $J$. If we assume that the charge
boson develops a gap and hence $<\cos (\sqrt{2\pi}\phi_c)>\neq 0$, then the
interaction in the spin sector reduces to the same one that occurs for
the spin ladder, Eq. (\ref{abelint}).  As argued above, this should gap
both spin excitations.  A different conclusion was reached in Ref.
\onlinecite{Fujimoto}, where it was claimed that $\phi_+$ does
not appear in the interaction and hence remains massless. This incorrect
conclusion was obtained because of a missing minus sign in the
transformation from nonabelian to abelian bosons, Eq. (\ref{gdef}).
The neccessity for the minus sign in the lower left matrix element can
be seen by observing that the constraint $\det g=1$ is not obeyed and
$\hbox{tr}(\vec \sigma g)$ is not purely anti-hermitean, without it.
(The ``note added in proof'' in Ref. \cite{Fujimoto}
reflects a realization of this error.\cite{Kawakami})

Away from half-filling, where the charge boson is massless, we may
analyze the spin sector much as for the zigzag chain.  In particular,
the Kondo interaction renormalizes to zero in the ferromagnetic case,
leaving all spin excitations gapless.  In the antiferromagnetic case, we
expect an exponentially small spin gap:
\begin{equation} \Delta \propto e^{-\hbox{constant}(v_F+v_s)/J}.
\end{equation}
Fujimoto and Kawakami\cite{Fujimoto} assumed instead that J
renormalized to some sort of strong coupling critical point
corresponding to vanishing spin gap for 1 branch of spin
excitations.  This assumption seems rather unlikely from the point of
view of the RG analysis, in light of the above comments, but was
motivated by physical considerations.  That is, it would seem that
somehow the left-moving spin excitations from the localized spins
interact with the right-moving spin excitations of the conduction
electrons (and vice versa) to form a gap.  While this seems reasonable at
half-filling, it becomes difficult to understand away from
half-filling.  If we consider the strong Kondo coupling limit then
localized spins form singlets with on-site conduction electrons. 
This clearly produces a gap at half-filling where there is one
conduction electron for each localized spin.  However, below
half-filling there is an excess of localized spins which may produce
gapless excitations.  In fact, for the ordinary Kondo lattice model
($K=0$) at strong coupling, $J >>t$, it has been shown by
Sigrist et al.\cite{Sigrist} that these left over spins form a
(gapless) ferromagnetic groundstate.  At weaker coupling Sigrist et
al. found a non-ferromagnetic phase whose properties were not very
well characterized.  Fujimoto et al.\cite{Fujimoto} assumed that this
phase corresponded to a single species of gapless spin excitations
(as well as gapless charge excitations).  

We do not find this argument
convincing and think that a spin-gap
phase may occur at weak coupling.  Evidence for this is
provided by our analysis of the zigzag spin chain in the previous
sections.  Note that in that case also a transition to a gapless phase
occurred for sufficiently large $J$, i.e. $J_1>J_{1c}\approx
4J_2$.  

A related phase with a 
spin gap but no charge gap has  been found numerically in the t-J
model away from half-filling, at $J$ of order $t$.  In this case, it is
apparently not associated with any spontaneous discrete symmetry
breaking, and may be thought of as a dimer fluid state.\cite{Ogata} 
The absence of spontaneous symmetry breaking is a consequence of the
gapless charge excitations.  The dimer order parameter also contains a
charge factor which has vanishing expectation value when the charge gap
vanishes.  

There is actually a limit of the generalized Kondo lattice model, below
half-filling, which is essentially equivalent to the t-J model: 
\begin{equation} J >>t, K>>t^2/J .\end{equation}  
The large J condition forces all conduction electrons to form singlets
with localized spins.  The unpaired localized spins can effectively hop
around via the hopping term. Their predominant interaction is the
Heisenberg term, $K$.  In addition they have various weak induced
interactions\cite{Sigrist} of $O(t^2/J )$.  These are the interactions
responsible for ferromagnetism in the pure Kondo lattice model at strong
coupling.  In the spin-gap phase of the t-J model these interactions
cannot change the behavior provided that they are small enough compared
to the gap.  This essentially constitutes a proof (given the numerical
results on the t-J model) that the generalized Kondo lattice model has a
spin-gap phase somewhere in its phase diagram away from half-filling.
How large a region of parameter space is in the spin-gap phase and
whether it includes the pure Kondo lattice model for some range of
doping and $J/t$ are open questions which we are investigating
numerically.\cite{Sikkema}

It was argued independently by Zachar et al. \cite{Zachar} that the
Kondo lattice should have a spin gap away from half-filling using a
different type of bosonization based on perturbing around a different
limit of the model where the Kondo interaction is strongly
anisotropic.  

\section{Conclusions}
The zigzag spin chain is gapless for weak ferromagnetic interchain coupling,
but has an exponentially small gap for small antiferromagnetic coupling.
This phase has a weak spontaneous dimerization, or broken translational
symmetry along with a finite-range incommensurate magnetic order.

Although we have presented results for a wide range of $J_2/J_1$,
our primary focus has been on the large $J_2/J_1$ region, where the
system is best thought of as two weakly coupled chains. Most
previous work has focused on smaller values of $J_2/J_1$.
Our results help to explain the behaviour of the quasi-one-dimensional
antiferromagnet, SrCuO$_2$, studied in Ref. \onlinecite{Matsuda}.
This compound is believed to be well described by the zigzag spin chain
(with very weakly coupled pairs of chains) with $J_2\approx 1000K$ and
$|J_2/J_1|$ in the range 10-1000.  The susceptibility appears to go to a
finite constant as $T\to 0$, apart from a low $T$ upturn attributed to
impurities.  This indicates the absence of a gap.  This is to be
expected from the results obtained here since the gap vanishes
exponentially with $J_2/J_1$ and should be competely negligible for this
range of couplings. Note that, if the interchain coupling had been of
ladder type rather than zigzag type, this gap would have been much
larger and perhaps observable in the available temperature range.  We
find that the gap is approximately $.42 J'$ where $J'$ is the
inter-chain coupling in the spin ladder, for $J'<<J$.  Thus the gap
might have been as large as 40K in the spin-ladder case and could have
shown up in the susceptibility measurements which went down to 1.7K.

The field theory description of the zigzag spin chain, in the limit,
$J_2>>J_1$ is closely related to the field theory description of the
decoupled spin sector in the doped generalized Kondo lattice in the
limit $K,t>>J$.  Our results on the zigzag spin chain suggest the
existence of a spin-gapped phase in the doped Kondo lattice.

\acknowledgements
I.A. thanks S. Coppersmith for interesting him in this problem and 
M.P.A. Fisher, A. Sikkema, H.J. Schulz. We thank D. Sen for 
pointing out an error in an early version of this paper.
S.R.W. acknowledges support from the Office of Naval Research under
grant No. N00014-91-J-1143, and
from the NSF under Grant No.\ DMR-9509945.
The research of I.A. was supported in part by NSERC of Canada.
Some of the calculations were performed at the
San Diego Supercomputer Center.


\begin{references}

\bibitem{Majumdar} C.K. Majumdar and D.K. Ghosh, J. Math. Phys. {\bf
10}, 1388 (1969).

\bibitem{Matsuda} M. Matsuda and K. Katsumata, J. Mag. Mag. Mat. {\bf
140-145}, 1671 (1995).

\bibitem{dagotto} E.\ Dagotto, J.\ Riera, and D.J.\ Scalapino, \prb
{\bf 45}, 5744 (1992).

\bibitem{strong} S.P.\ Strong, and A.J.\ Millis, \prl {\bf 69}, 2419 (1992).

\bibitem{barnes} T.\ Barnes et al., \prb {\bf 47}, 3196 (1993).

\bibitem{rice} T.M. Rice, S. Gopalan, and M. Sigrist, Europhys. Lett.
{\bf 23}, 445 (1993);  M.\ Sigrist, T.M.\ Rice, and F.C.\ Zhang, 
{\sl Phys. Rev. } B {\bf 49}, 12058 (1994).\par

\bibitem{rvbprl} S.R.\ White, R.M.\ Noack, and D.J.\ Scalapino,
\prl {\bf 73}, 886 (1994).

\bibitem{Schulz1} H.J. Schulz, Phys. Rev. {\bf B53}, 2959 (1996).

\bibitem{Balents} L. Balents and M.P.A. Fisher, UCSB preprint NSF-ITP-9523,
cond-mat/9503045.

\bibitem{dmrg} S.R. White, \prl {\bf 69}, 2863 (1992),
\prb {\bf 48}, 10345 (1993).

\bibitem{tonegawa} T. Tonegawa and I. Harada, 
{\sl J. Phys. Soc. Jpn. } {\bf 56}, 2153 (1987).

\bibitem{bursill}  R.J. Bursill , G.A. Gehring, 
D.J.J. Farnell, J.B. Parkinson, Chen Zeng, T. Xiang, preprint, 
cond-mat/9511044.

\bibitem{chitra} R. Chitra, S. Pati, H.R. Krishnamurthy, D. Sen, and
S. Ramasesha, \prb {\bf 52}, 6581 (1995).

\bibitem{Haldane} R. Julien and F.D.M. Haldane, Bull. Am. Phys. Soc.
{\bf 28}, 34 (1983).

\bibitem{Affleck1}I. Affleck, {\it Fields, Strings
and Critical Phenomena}, ed. E. Br\'ezin and J. Zinn-Justin
(North-Holland, Amsterdam, 1990), 563.

\bibitem{Eggert} S. Eggert and I. Affleck, Phys. Rev. {\bf B46}, 10866
(1992).

\bibitem{Eggert2} S. Eggert, preprint cond-mat/9602026.

\bibitem{okamoto} K. Okamoto and K. Nomura, Physics Letters A {\bf
169}, 433 (1992).

\bibitem{Affleck} I. Affleck, T. Kennedy, E. Lieb and H. Tasaki,
Comm. Math. Phys. \underbar{115}, 477 (1988).

\bibitem{Schulz2} H.J. Schulz, Phys. Rev. \underbar{B34}, 6372 (1986).

\bibitem{Strong} S.P. Strong and A.J. Millis, Phys. Rev. Lett. \underbar
{69}, 2419 (1992).

\bibitem{Shelton} D. G. Shelton, A. A. Nersesyan and A. M. Tsvelik,
preprint, cond-mat/9508047.

\bibitem{georgia} R.M.\ Noack, S.R.\ White and D.J.\
Scalapino, in {\it Computer Simulations in Condensed Matter Physics
VII}, Eds. D.P. Landau, K.K. Mon, and H.B. Sch\"uttler (Springer Verlag,
Heidelberg, Berlin, 1994), p.\ 85-98.

\bibitem{improvedmrg} S.R. White, to be published.

\bibitem{boseconden} E. Sorensen and I. Affleck,
Phys. Rev. Lett. {\bf 71}, 1633 (1993).

\bibitem{Ogawa} M.Y. Ogawa et al. Phys. Rev. Lett. {\bf 57},
1177 (1986); Phys. Rev. {\bf B39}, 10682 (1989);
J. Am. Chem. Soc {\bf 109}, 1115 (1987).

\bibitem {Quirion} G. Quirion et al. Phys. Rev. {\bf B37},
4272 (1988).

\bibitem{Fujimoto} S. Fujimoto and N. Kawakami, J. Phys.
Soc. Japan, \underbar{63}, 4322 (1994).

\bibitem{Kawakami} N. Kawakami, private communication.

\bibitem{Sigrist} M Sigrist, H Tsunetsugu, K Ueda and TM Rice, Phys.
Rev. \underbar{B46}, 13838 (1992).

\bibitem{Ogata} M. Ogata, M. Luchini, S. Sorella and F. Assaad, Phys.
Rev. Lett. \underbar{66}, 2388 (1991); C. Hellberg and E. Mele, Phys.
Rev. \underbar{B48}, 646 (1993). 

\bibitem{Sikkema} A. Sikkema, S.R. White and I. Affleck, in progress.

\bibitem{Zachar} O. Zachar, S.A. Kivelson and V.J. Emery, preprint,
cond-mat/9508109.


\end{references}
\end{document}